\documentclass[twocolumn,10pt,prl, aps,superscriptaddress,showpacs]{revtex4-1}

\usepackage[colorinlistoftodos]{todonotes}
\usepackage[utf8]{inputenc}

\usepackage{subfigure}

\usepackage{dcolumn}
\usepackage{amsmath,amssymb}
\usepackage{bm}
\usepackage{color}
\usepackage{latexsym}
\usepackage{psfrag,graphicx}
\usepackage{color}
\usepackage[english]{babel}
\usepackage{latexsym}
\usepackage{epsf} 
\usepackage{amsfonts}
\usepackage{natbib}
\usepackage{epstopdf}
\usepackage{appendix}
\usepackage{changes}
\usepackage{siunitx}
\usepackage{gensymb}

\usepackage{hyperref}
\usepackage[capitalize]{cleveref}
\hypersetup{
    colorlinks,
    citecolor=blue,    filecolor=blue,
    linkcolor=blue,    urlcolor=blue
}

\newcommand{\WIS}{Department of Physics of Complex Systems, Weizmann Institute of Science, Rehovot 76100, Israel}
\newcommand{\LOA}{Laboratoire d’Optique Appliqu\'ee, ENSTA Paris, CNRS, Ecole Polytechnique, Institut Polytechnique de Paris, 828 Bd des Mar\'echaux, 91762 Palaiseau, France}
\newcommand{\Ardop}{Ardop Engineering, Cit\'e de la Photonique, 11 Avenue de la Canteranne, B\^at. Pl\'eione, 33600 Pessac, France}

\begin{document}

\title{Low divergence proton beams from a laser-plasma accelerator at kHz repetition rate}

\author{Dan~Levy}\email{dan.levy@weizmann.ac.il}\affiliation{\WIS}
\author{Igor~A.~Andriyash}\email{igor.andriyash@ensta-paris.fr}\affiliation{\LOA}
\author{Stefan~Haessler}\email{stefan.haessler@cnrs.fr}\affiliation{\LOA}
\author{Jaismeen~Kaur}\affiliation{\LOA}
\author{Marie~Ouill\'e}\affiliation{\LOA}\affiliation{\Ardop}
\author{Alessandro~Flacco}\affiliation{\LOA}
\author{Eyal~Kroupp}\affiliation{\WIS}
\author{Victor~Malka}\affiliation{\WIS}
\author{Rodrigo~Lopez-Martens}\affiliation{\LOA}

\begin{abstract}
Proton beams with up to $100$~pC bunch charge, $0.48$~MeV cut-off energy and divergence as low as a $3\degree$ were generated from solid targets at kHz repetition rate by a few-mJ femtosecond laser under controlled plasma conditions. The beam spatial profile was measured using a small aperture scanning time-of-flight detector. Detailed parametric studies were performed by varying the surface plasma scale length from $8$ to $80$~nm and the laser pulse duration from $4$~fs to $1.5$~ps. Numerical simulations are in good agreement with observations and, together with an in-depth theoretical analysis of the acceleration mechanism, indicate that high repetition rate femtosecond laser technology could be used to produce few-MeV protons beams for applications.
\end{abstract}
                                          
\maketitle

Over the past two decades, ion acceleration with intense femtosecond laser pulses has emerged as a promising alternative to conventional accelerators due to its much higher instantaneous flux and potentially lower costs and smaller facility size. Laser-accelerated protons are already used for target studies for fast ignition  \cite{roth_fast_2001}, studies of warm dense matter \cite{patel_isochoric_2003} and probing fast phenomena in laser matter interaction \cite{borghesi_electric_2002}. However, societal applications of laser accelerated protons such as proton therapy \cite{malka_principles_2008} and nuclear physics research \cite{ledingham_applications_2003} require source parameters that are still out of reach for existing laser-based accelerators \cite{lefebvre_numerical_2006, linz_laser-driven_2016}. Intense research efforts are dedicated to improving the source performance in terms of particle energy, average flux, beam brightness and overall source efficiency. 

Ion acceleration with ultraintense lasers (${I_\text{las}\gtrsim 10^{18}\,\rm W/cm^2}$) is currently mostly realized by irradiating thin ($\sim0.1-10\,\rm\mu m$) foils, utilizing the well-known target normal sheath acceleration (TNSA) mechanism \cite{wilks_energetic_2001, ceccotti_proton_2007} which is driven by the thermal pressure of hot electrons generated by the laser. Commercially available $100$~TW-class lasers deliver TNSA proton beams with energies reaching up to few tens of MeV and $10^{11}-10^{13}$ protons per shot above $1$~MeV, depending on target and laser parameters \cite{borghesi_ion_2019}. These lasers typically operate at $1-10$~Hz, yet a continuous operation with thin foils at this rate is very challenging due to target replenishing needs and debris contamination in the vacuum chamber.

Another major shortcoming of TNSA proton beams compared to the ones from conventional accelerators is their much larger divergence, which ranges from $10\degree$ to $30\degree$ FWHM \cite{snavely_intense_2000,cowan_ultralow_2004,brambrink_transverse_2006}. The demand for high brightness has driven considerable research efforts aiming at reducing the angular spread of laser-generated beams. Such efforts include additional collimation devices \cite{toncian_ultrafast_2006, kar_dynamic_2008, kar_guided_2016}, special target geometries \cite{sonobe_suppression_2005,miyazaki_high-energy_2005,nakamura_robustness_2007,yu_quasimonoenergetic_2009, pae_proposed_2009-1, ma_high-quality_2009, chen_enhanced_2009,chen_target_2010,wilks_energetic_2001, okada_energetic_2006, liu_energetic_2008, ali_bake_energetic_2013, yasen_enhancement_2019, matys_laser-driven_2020,klimo_monoenergetic_2008-1} and targets which enable other acceleration mechanisms \cite{robinson_radiation_2008-2, qiao_stable_2009, zhuo_quasimonoenergetic_2010,park_ion_2019,pae_low-divergence_2020-1}. In many cases, target fabrication and handling, together with stringent laser contrast requirements for laser-solid acceleration, hinder the use of these targets in applications where a stable operation at a high repetition rate is necessary.

\begin{figure}
\includegraphics[width=0.8\linewidth]{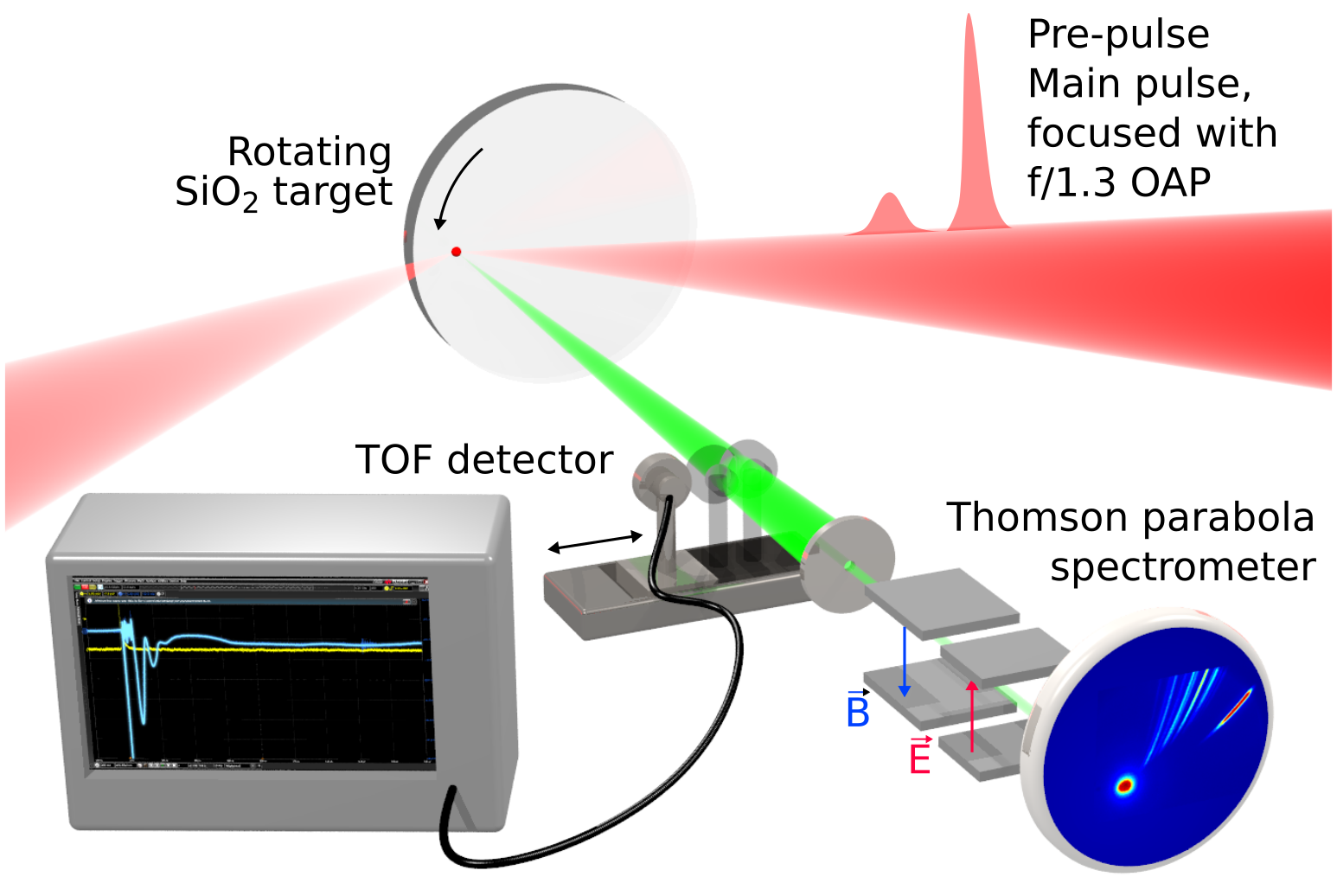}
\caption{Experimental setup.}
\label{fig:experimental_setup}
\end{figure}

With the latest advancements in laser technology, ion acceleration using kHz lasers is emerging as a promising path for generating proton beams of sufficient flux and energy. Studies include both solid targets \cite{hou_mev_2009,veltcheva_brunel-dominated_2012} as well as other types. Recently, Morrison \textit{et al.} demonstrated kHz proton beams of up to 2 MeV using a mJ-class laser and a thin liquid sheet target \cite{morrison_mev_2018}. Beam divergence with this target remains however similar to typical TNSA beams, and thus remains a serious limitation for practical use.

\begin{figure*}
\includegraphics[width=0.8\textwidth]{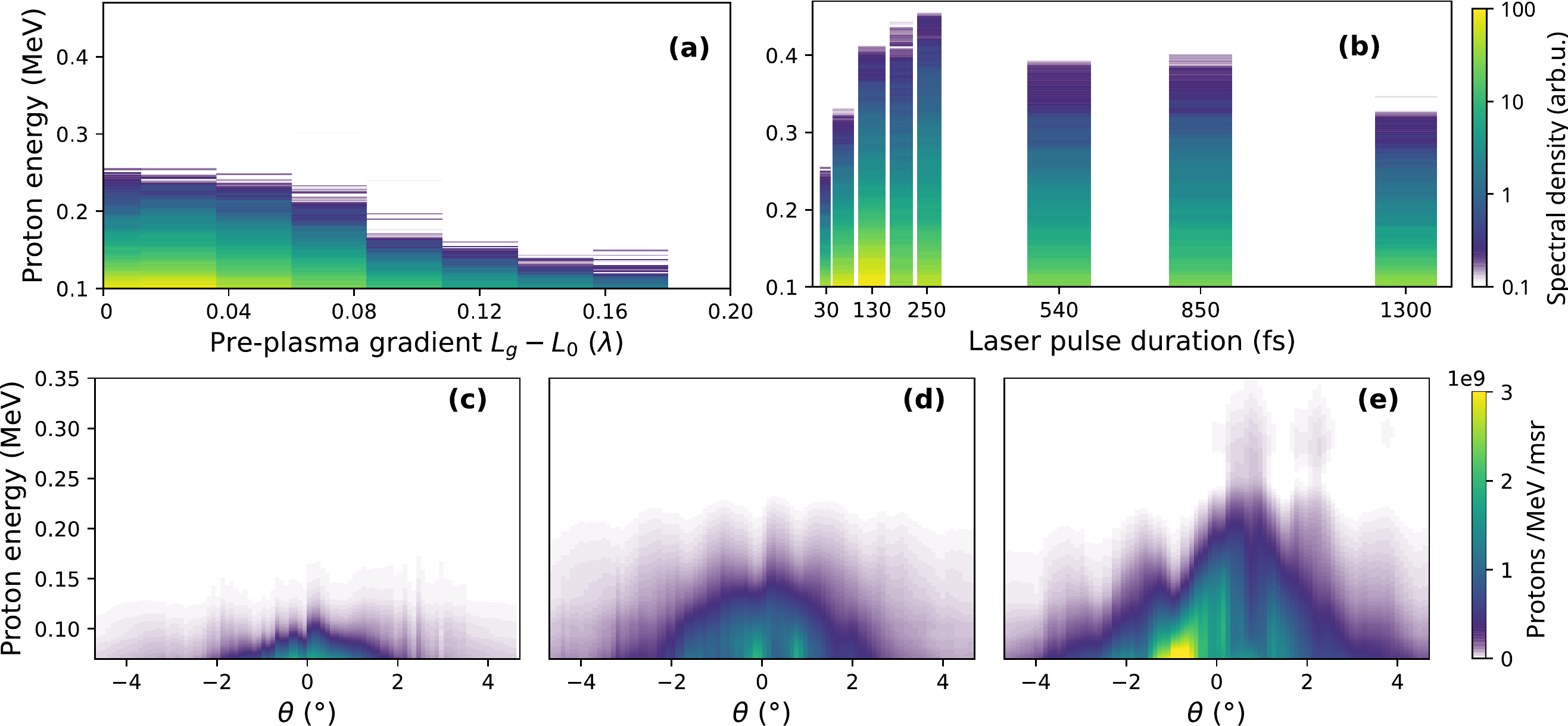}
\caption{TPS proton spectra in arbitrary units and logarithmic scale, for $\tau_\text{las}=27$~fs and varying plasma scale length (a), and for 0 prepulse delay and varying pulse duration (b). Angular-spectral proton number distributions measured with the calibrated TOF detector for $4$~fs (c), $27$~fs (d) and $200$~fs (e).}
\label{fig:TOF_spectra_colormaps}
\end{figure*}

In this Letter we demonstrate and analyze a robust, efficient and easy-to-implement method for delivering highly collimated proton beams at kHz repetition rate using few-mJ femtosecond pulses. The underlying mechanism is essentially different from TNSA, and its previous studies suggested that it is associated with electron motion in the Brunel heating process \cite{veltcheva_brunel-dominated_2012}. We revisit this process experimentally with vastly improved laser performance in terms of intensity and temporal contrast. In addition, we measure for the first time the angular and energy distributions of the accelerated protons over a wide range of interaction conditions by varying both pulse duration and plasma gradient length. The obtained data was used to construct a comprehensive theoretical description supported by the detailed numerical modeling. Our findings give a clear physical picture of the acceleration and its scalability for applications.

The experiment was carried out at the Salle Noire facility of Laboratoire d'Optique Appliqu\'ee (LOA). The laser delivers $27$~fs pulses at 1 kHz repetition rate with high temporal contrast (${> 10^{10}}$ at $10$~ps). The pulse duration can be reduced to $4$~fs in a hollow-fiber compressor \cite{ouille_relativistic-intensity_2020} and increased up to $1500$~fs by adding intra-chain group delay dispersion (GDD), while keeping the same energy and spatial intensity distribution on target. The peak intensity on target can thus be tuned from $10^{16}$ to $10^{19}\,\rm W/cm^2$. A schematic layout of the experiment is shown in \cref{fig:experimental_setup}, where the $2.5$~mJ, p-polarised pulses are focused by an $f/1.3,\,30\degree$ off-axis parabolic mirror (OAP) down to a $R_\text{las}\approx 1.8$~$\mu$m FWHM focal spot at $55\degree$ incidence angle at the surface of a rotating fused silica optical substrate. A time-delayed pre-pulse is added by picking off $\approx 4\%$ of the main pulse through a holey mirror, and focusing it with the same OAP to a larger $13\, \rm\mu m$ FWHM spot. The preplasma gradient scale length $L_\text{g} \approx L_0 + c_s t_d$ can be controlled by changing the relative delay $t_d$ between the pre-pulse and the main pulse, where $L_0$ corresponds to an additional expansion induced by the main pulse pedestal due to its finite temporal contrast. The plasma expansion velocity $c_s$ is measured using spatial domain interferometry (SDI) \cite{bocoum_spatial-domain_2015}. The high-stability rotating target holder~\cite{borot_high_2014} keeps the target surface position stable within the few-micron Rayleigh length of the tightly focused main pulse, while refreshing the target surface for shots at $1\,\rm kHz$.

The proton energy and angular distributions were measured with a Thomson parabola spectrometer (TPS) and a charge-calibrated scanning time-of-flight (TOF) detector. The TPS accepts ions in the normal direction through a $300\,\rm \mu m$ pinhole placed at about $0.6$~m from the target. The TOF detector consists of a $6$~mm~diameter microchannel plate (MCP) placed $375$~mm away from the irradiated point and connected to a $8$~GHz oscilloscope. This setup provides an acceptance angle of $0.9\degree$, such that the broadening of the measured angular profile due to convolution with the detector size is negligible. To the best of our knowledge, this is the first time a TOF detector was used to measure the angular distribution of the beam. Further details regarding both detectors are available in the Supplemental Material \cite{SM}.

Figures~\ref{fig:TOF_spectra_colormaps}(a-b) show proton energy spectra acquired with the TPS, where each spectrum is an average of 100 consecutive shots. In \cref{fig:TOF_spectra_colormaps}(a) the pulse duration is fixed at $\tau_\text{las}=27$~fs and the pre-pulse delay is varied. The highest proton energies around $0.25$~MeV are achieved for the steepest gradient and drop below the detection threshold of $0.1$~MeV for the gradients $L_\text{g}\gtrsim \lambda/6$. In \cref{fig:TOF_spectra_colormaps}(b) the gradient is kept the steepest ($0$~ps delay) and the pulse duration is varied. We observe an optimum around $100-300$~fs, where the maximum proton energy $W_\text{max}$ reaches $0.48\pm0.02$~MeV. For the longer pulses, the maximum proton energy slowly decreases.

Figures~\ref{fig:TOF_spectra_colormaps}(c-e) show the angle-resolved TOF spectra measured for the sharpest plasma gradient and for driving pulse durations of $4$~fs, $27$~fs and $200$~fs. For each angle, each spectrum shown is an average of 100 consecutive shots. All three measurements in \cref{fig:TOF_spectra_colormaps}(c-e) show similar angular profiles and in all cases the energy-integrated divergence is $\approx 3\degree$~(FWHM). Assuming a two-dimensional Gaussian angular profile, we calculate the total charge (above $0.1$~MeV) to be $12\pm2.4$~pC, $50\pm10$~pC and $98\pm20$~pC, respectively. The single-axis measurement is indeed fitted very well by a Gaussian and we expect the beam asymmetry to be small due to the symmetry of the system, as was observed in \cite{hou_mev_2009} (albeit with much larger beam divergence). Our estimation of $20\%$ error originates from uncertainties in the angular distribution and the TOF calibration, as described in the Supplemental Material \cite{SM}.  

\begin{figure*}
\includegraphics[width=0.95\linewidth]{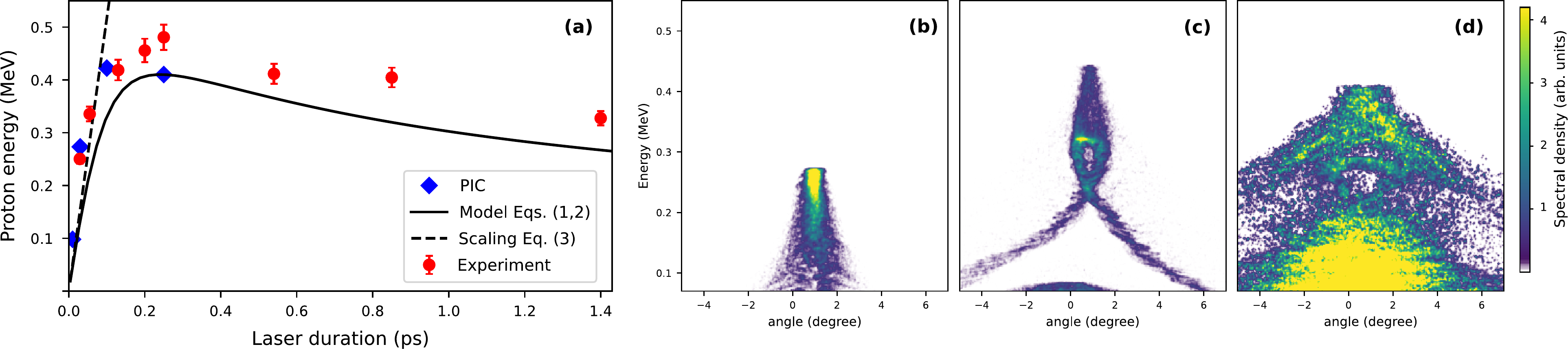}
\caption{(a) Maximum proton energy as a function of laser pulse duration obtained experimentally (red dots), in PIC simulations (blue diamonds), from numerical integration of proton motion with \cref{Eq:Z0FullField,Eq:Eacc} (black solid line), and the scaling \cref{Eq:LinScale} (dashed line). (b-d) Final angular spectral distributions of protons for $30$~fs (b), $100$~fs (c) and $250$~fs (d) laser pulse durations.}
\label{fig:PIC_scan}
\end{figure*}

\Cref{fig:TOF_spectra_colormaps} points out that the acceleration benefits from the shortest, laser-contrast-limited plasma gradient, and that the highest cut-off energies are reached for $\tau_\text{las} \sim 100-200$~fs, at the point when the growth $W_\text{max}\propto\tau_\text{las}$ switches to a slow decrease $W_\text{max}\propto I_\text{las} \propto 1/\tau_\text{las}$. This first regime suggests that for the short pulses, the acceleration is directly driven by the laser-plasma interaction, while not strongly depending on the laser intensity. Let us consider a physical picture of a solid plasma slab irradiated obliquely by a short sub-relativistic laser pulse, with field strength ${a_0 = e E_\text{las} \lambda / 2\pi m_\text{e} c^2 < 1}$, where $E_\text{las}=\sqrt{2 I_\text{las}/\epsilon_0 c}$ is the peak electric field, $\lambda$ is its wavelength, $e$ and $m_\text{e}$ are the electron charge and mass, respectively, and $c$ is the speed of light in vacuum. The laser field penetrates the preplasma until it reaches the reflective layer with electron density $n_\text{pe}\gtrsim n_\text{c}$, where ${n_\text{c} = \pi/r_\text{e}\lambda^2}$ is the critical plasma density and $r_\text{e}$ is the classical electron radius. Within this layer, the laser introduces the radiation pressure ${P_\text{rad}=2\cos^2\!\theta\, I_\text{las}/c}$, which expels plasma electrons until it becomes balanced by the electrostatic pressure $P_\text{es}$. For a short preplasma gradient, $L_\text{g}\ll\lambda$, the laser and electrostatic fields beyond the penetrated layer are instantly screened by the dense plasma, enabling the well-known Brunel absorption \cite{brunel_not-so-resonant_1987}. Owing to this screening, electrons escape deep into the plasma without building up a dense negative layer, thus allowing ions to generate a long-range accelerating field. For the considered laser intensities, the preplasma consists of only partially ionized Si, O and C ions, and a small fraction of protons (see Sec.~I.C and S-Fig.~3 in the Supplemental Material \cite{SM}). The latter have a much higher charge-to-mass ratio than heavier ions, ${1/m_\text{p} \gg Z_i/M_i}$, and can thus be quickly accelerated in the electrostatic field of this positively charged preplasma before its own Coulomb explosion.

Let us now consider a simple quantitative description of the physical process described above. We assume a plasma with initial density, ${n_\text{p0}\gg\, n_\text{c}}$, that is uniform for $z<0$, and falls as ${n_\text{p}=n_\text{p0} \exp(-z/L_\text{g})}$ for $z>0$. The thickness of the un-neutralized ion layer created by the laser pressure can be estimated from the balance condition, $P_\text{es}\simeq P_\text{rad}$, as:
\begin{equation}\label{Eq:Z0FullField}
 z_0 \simeq L_\text{g} \ln\! \left(\cfrac{e  n_\text{p0} L_\text{g}}{2\cos\theta} \sqrt{\frac{c}{\epsilon_0 I_\text{las}}} \,\right)\,,
\end{equation}
where $\epsilon_0$ is the vacuum permittivity (for details see \cite{SM}). For $z>z_0$, the field profile can be estimated theoretically by considering the model of a thin charged disk:
\begin{align}\label{Eq:Eacc}
  E_\text{z}(t, z) \,=\, & \cfrac{\hat{E}_\text{las}(t)\,\cos\theta}{2} \; 
 \left[1-\exp\left(-\cfrac{z-z_0}{L_\text{g}}\right) \right]\times\nonumber \\
 &\times\left(1-\cfrac{z}{\sqrt{z^2+R_\text{i}^2}}\right)\,,
\end{align}
where $\hat{E}_\text{las}$ is the amplitude of laser electric field, and the size of the charged ion disk is determined by the projected laser spot size as $R_\text{i}=R_\text{las}/\cos\theta$. In the particular limit when the acceleration length is short compared to the field's scale $R_\text{i}$ and long compared to the preplasma scale $L_\text{g}$, one may discard the geometric factors in \cref{Eq:Eacc} and calculate the proton energy obtained in the field ${E_\text{z}=\hat{E}_\text{las}(t)\,\cos\theta/2}$ over the laser duration as:
\begin{equation}\label{Eq:LinScale}
 W_\text{max} = \sqrt{\frac{\ln 2}{\pi}}\; \cfrac{e^2}{m_p c \,\epsilon_0}\;
    \cfrac{W_\text{las} \,\tau_\text{las}}{R_\text{i}^2} \,,
\end{equation}
where $W_\text{las}$ is the laser energy. Noting that the field in \cref{Eq:Eacc} vanishes for $z\gtrsim R_\text{i}/2$, and limiting the full interaction time by $2 \tau_\text{las}$, we can define the validity condition for \cref{Eq:LinScale} as $\langle v_p\rangle \tau_\text{las} \lesssim R_\text{i}/4 $, where $\langle v_p\rangle=\sqrt{W_\text{max}/2m_p}$ is the proton velocity averaged over the acceleration duration. At its limit, this condition defines a case when protons reach the vanishing field by the end of the interaction, and it thus defines the optimal acceleration regime. Applying this condition to \cref{Eq:LinScale}, we obtain the scaling:
\begin{equation}\label{Eq:OptScale}
 W_\text{opt} = \left(\sqrt{\frac{2 \ln 2}{\pi m_p}} \cfrac{e^2}{4 c \,\epsilon_0} \cfrac{W_\text{las}}{R_\text{i}}\right)^{2/3}\,,
\end{equation}
which suggests that the maximum cut-off energy depends neither on the plasma nor on the preplasma features (as long as $L_\text{g}\ll\lambda$), but is determined solely by the laser energy and its projected spot size. More details on the theoretical analysis can be found in \cite{SM}.

The described mechanism relies on the persisting $I_\text{las}$, and as soon as the laser pressure is gone, electrons flow back into the charge separation region and can drive a consequent TNSA process. For a short preplasma and a short laser, the electrons that refill that region are cold, and this remnant acceleration turns out to be insignificant. For the longer laser pulses, the expansion of the heavier ions breaks the sharp plasma interface, and TNSA occurs during the laser action. This can also be observed in \cref{fig:TOF_spectra_colormaps}(c-e), where the short-pulse cases (c,d) show all energies being contained within $2\degree$ half-angle, while in the 200~fs laser case \cref{fig:TOF_spectra_colormaps}(e), the angular divergence of protons with $W_p<0.15$~MeV increases. We note that the TOF measurements in the latter case also indicate that low energy protons (below 0.07 MeV) get intermixed with heavier ions, resulting in an increased divergence of the aggregate signal (up to $5\degree$ FWHM). This divergence degradation at longer interaction times cannot be easily avoided, and it restricts the optimal laser durations to not exceed 100-200~fs (weakly depending on the ion composition).

For more details and to verify this interpretation we consider 2D particle-in-cell (PIC) simulations using the code WarpX \cite{vay_modeling_2021}. In the simulations, the pre-plasma has an exponential profile with $L_\text{g}=8$~nm, and the laser pulse of $3$~mJ energy is focused on the surface into a $1.8$~$\mu$m spot (see \cite{SM} for details). \Cref{fig:PIC_scan}(a) compares the proton cut-off energy for various pulse durations measured in the experiment, simulations and provided by the analytical model \cref{Eq:Z0FullField,Eq:Eacc,Eq:LinScale}. All data sets agree well with each other. While the cut-off energy follows the measured and predicted flat scalings for long laser pulses, the details of the protons' spectral-angular distributions in \cref{fig:PIC_scan}(b-d) reveal clear signatures of the TNSA process which severely degrades the resulting divergence. For $\tau_\text{las} = 30$~fs, all protons remain well collimated, $\tau_\text{las} = 100$~fs develops high divergence for slower protons, and $\tau_\text{las} = 250$~fs generates a divergent TNSA beam. Note that this TNSA signature is much more pronounced in the modeling than in the experimental measurements, which we may explain by the specificity of the simulated 2D geometry. While the considered proton acceleration is essentially one-dimensional and should not depend significantly on the transverse dimensions, the 2D PIC simulations are known to greatly enhance the TNSA process, and were shown to result in twice more efficient acceleration compared to the full 3D geometry \cite{dhumieres_optimization_2013}. 

\begin{figure}
\includegraphics[width=0.75\linewidth]{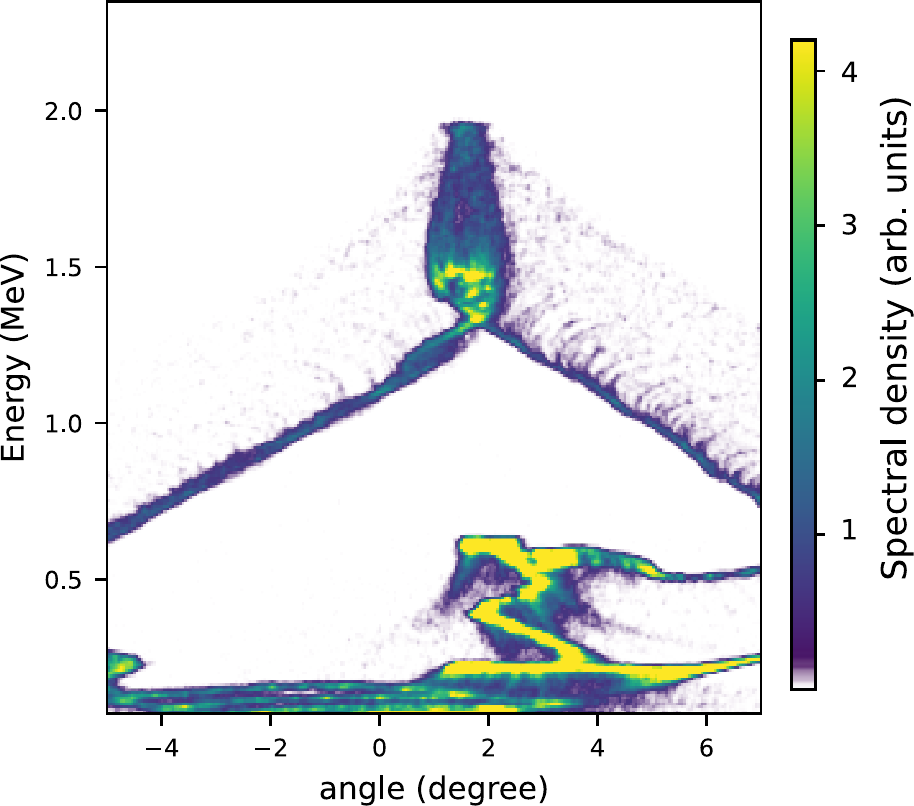}
\caption{ Final angular-spectral distribution of protons for laser energy of 30~mJ and 70~fs pulse duration.}
\label{fig:PIC_30mJ}
\end{figure}

We may refer to the described mechanism as radiation pressure assisted Coulomb explosion (RPACE). So far we have obtained the scaling of the maximum proton energies and have identified optimal conditions, but have also discovered that subsequent TNSA degrades the beam quality, thus limiting RPACE to the shorter laser durations. This process is governed by multiple parameters (laser intensity, duration, ion composition), and one may expect that the rate of such degradation increases for higher laser energies as the electron temperature grows and TNSA intensifies. This tightens the limitation of RPACE with respect to the laser duration and imposes a sub-optimal source performance with the maximum proton energy following the scaling \cref{Eq:LinScale}. Practically, for a given laser energy and focusing geometry, the maximal pulse duration that preserves the low proton beam divergence should be found. In order to check the energy scaling, we consider a case with a laser of 30~mJ energy, 70~fs pulse duration, and 2.5~$\mu$m focal spot size. The simulated angular-spectral protons distribution is shown in \cref{fig:PIC_30mJ} and indicates the energy cutoff at $\approx 2$~MeV, which is in a very good agreement with the expected cutoff at $1.9$~MeV according to \cref{Eq:LinScale}. While faster particles remain well-collimated, \cref{fig:PIC_30mJ} indicates that the protons with energies $W_p<1.4$~MeV suffer the TNSA-induced increased divergence. We anticipate that in the full 3D geometry this threshold should be well below 1~MeV. In \cref{fig:PIC_30mJ}, one may also note the angular modulations appearing for $W_p<0.5$~MeV. These result from the surface plasma wave, which is excited by the laser and modulates the transverse spatial structure of the accelerating fields.

In summary, we have measured proton beams with unprecedented low divergence, generated at kHz repetition rate over a wide range of laser and plasma parameters. The divergence was found to be as low as $3\degree$ FWHM, about an order of magnitude smaller than thin foil TNSA proton beams. Proton energies reaching up to $0.48\pm0.02$~MeV for an optimal pulse duration ranging from $150\,\rm fs$ to $300\,\rm fs$ were measured with the steepest plasma-vacuum interface. The total charge reached a maximum of $98\pm20$~pC above $0.1$~MeV, giving an average current of $\sim0.1\,\rm \mu A$. The RPACE acceleration mechanism was identified in simulations and theory, showing that the surface protons Coulomb-explode after the laser radiation pressure evacuates preplasma electrons into the dense target. The scaling and optimal conditions for proton energies were obtained and discussed.

The demonstration of these unique proton beams is promising for numerous applications, some of which are within reach. Notable such applications are ion implantation \cite{lorusso_modification_2005, torrisi_unconventional_2016} and the production of radioisotopes for positron-emission tomography (PET), which requires energies of a few MeV~\cite{fritzler_proton_2003, clarke_production_2013}. Such energies could be reached with readily available systems providing few-tens-mJ pulses according to our analysis.

\begin{acknowledgments}
This project has received funding from the European Union’s Horizon 2020 research and innovation program under grant agreements 871124 and 694596 (LaserLab-Europe) and  694596 (ERC Advanced Grant ExCoMet), as well as from the Agence Nationale de la Recherche (ANR) under grant agreement ANR-10-LABX-0039-PALM. This work was performed using HPC resources from GENCI–TGCC (Grant 2021-A0110510062). This research used the open-source particle-in-cell code WarpX \url{https://github.com/ECP-WarpX/WarpX}, primarily funded by the US DOE Exascale Computing Project. We acknowledge all WarpX contributors.
\end{acknowledgments}

\bibliography{bibliography.bib}
\bibliographystyle{apsrev4-1}

\end{document}


\title{Supplemental Material: \\Low divergence proton beams from a laser-plasma accelerator at kHz repetition rate}

\author{Dan Levy}\email{dan.levy@weizmann.ac.il}\affiliation{\WIS}
\author{Igor A. Andriyash}\email{igor.andriyash@ensta-paris.fr}\affiliation{\LOA}
\author{Stefan Haessler}\email{stefan.haessler@cnrs.fr}\affiliation{\LOA}
\author{Marie Ouill\'e}\affiliation{\LOA}\affiliation{\Ardop}
\author{Jaismeen Kaur}\affiliation{\LOA}
\author{Alessandro Flacco}\affiliation{\LOA}
\author{Eyal Kroupp}\affiliation{\WIS}
\author{Victor Malka}\affiliation{\WIS}
\author{Rodrigo Lopez-Martens}\affiliation{\LOA}
                                          
\maketitle

\section{proton diagnostics in the experiment}

\subsection{TOF measurement}\label{sec:TOF_meas}

The TOF detector was mounted on a translation stage which moved in parallel to the rotating fused silica slab target. For each stage position, the signals of 100 consecutive shots were recorded by the oscilloscope. The angular profiles in different conditions were measured by moving the stage in steps of $0.65$~mm, corresponding to about $0.1\degree$.

\begin{figure}[!h]
\includegraphics[width=0.7 \linewidth]{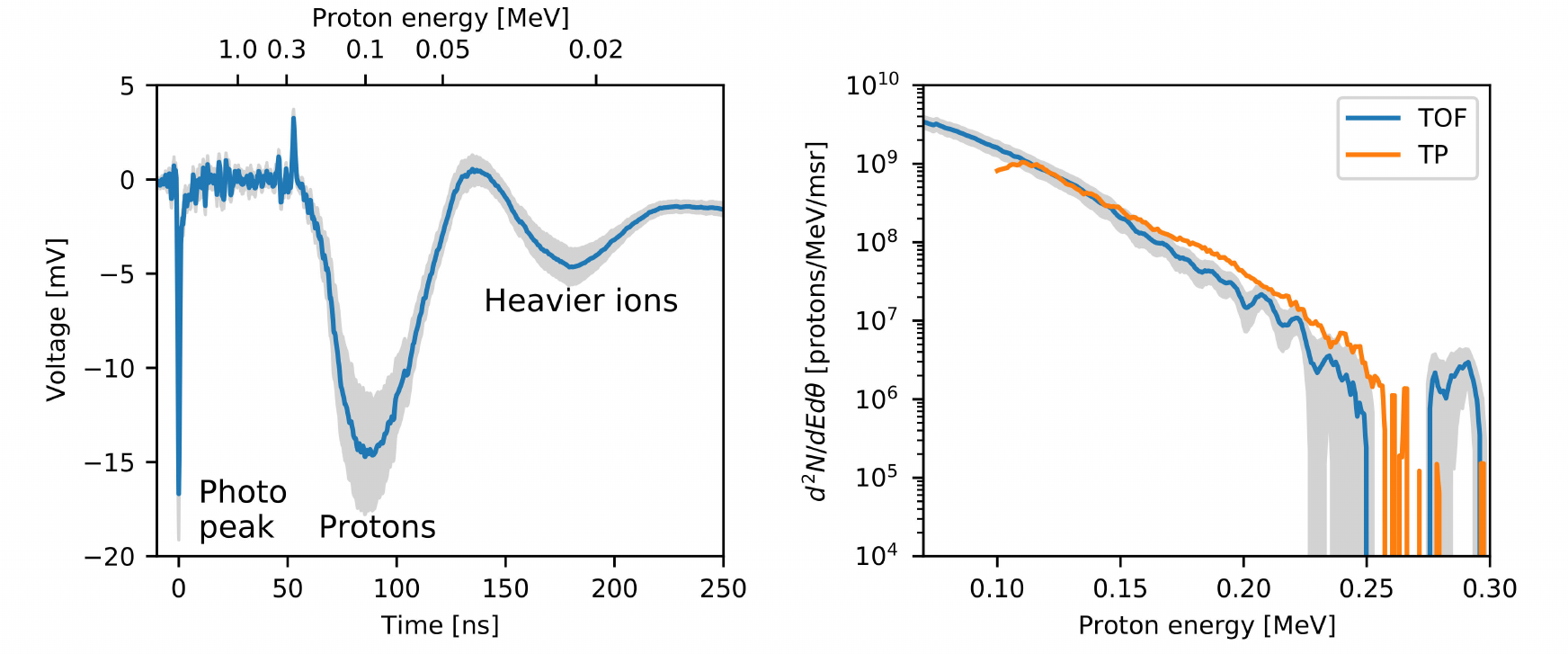}
\caption{(a) Typical TOF signal measurement. (b)}
\label{fig:TOF_signal_spect}
\end{figure}

A typical TOF signal is shown in \cref{fig:TOF_signal_spect}(a). These measurements were taken for a pulse duration of $27$~fs, no prepulse delay ($0$~ps) and at the target normal direction ($0\degree$). The curve is an average of the 100 consecutive shots, smoothed by a rolling average over the response time of the detector ($1\,\rm ns$). The shaded gray area shows the standard deviation of the shots. The first sharp peak corresponds to the electromagnetic radiation emitted from the laser-plasma interaction and defines the time $t=0$. The first protons start to arrive at $t\approx 50$~ns, with a peak signal at $t\approx 90$~ns, followed by the heavier ions with a signal peaked at $t\approx 180$~ns. The noise peak at $51\:$ns/$\approx0.3$~MeV was ever-present in all our measurements and is due to electrical noise in the circuit. At this energy and above, the spectrum is more reliably measured with the Thomson parabola detector.

In \cref{fig:TOF_signal_spect}(b) we show the proton energy spectrum extracted from the signal in \cref{fig:TOF_signal_spect}(a) alongside the spectrum extracted from the Thomson parabola spectrometer measured under the same conditions. The TPS spectrum is also an average of 100 consecutive shots. Since the TOF detector was charge calibrated (details below) and the TPS was not, the TOF spectrum is given in units of protons/MeV/msr, while the TPS spectrum is in arbitrary units and is juxtaposed for comparison with the TOF signal. At energies beyond $0.11$~MeV the two curves follow together fairly closely, showing a cutoff energy at about $0.25$~MeV followed by noise. The decline in the TPS spectrum below $0.11$~MeV is due to the signal reaching close to the edge of the MCP. 

The lower energy limit of proton detection is determined by the simultaneous arrival of heavier ions to the detector. In  \cref{fig:TOF_signal_spect}(a) the two signals are separated, which is generally not the case as shown in \cref{fig:TOF_mixing}(a). We find that 0.07 MeV are a good separator for all shots and choose it as a lower bound. Protons of lesser energy are expected to diverge more as they experience a longer time of interaction with transverse forces. Upon examination of the aggregate signal of both low energy protons and high energy heavier ions we see that the divergence remains small ($\approx5\degree$), so the overall proton beam divergence is not significantly affected.

\subsection{TOF calibration}
Determining the total emitted proton charge per shot from the TOF time-dependent signal requires calibration of the detector. This TOF detector is based on a $\sim 0.5$~mm thick MCP, grounded on the front side (facing the incoming particles), and biased by a positive voltage ($600$~V) at the opposite anode side using a bias tee circuit. We have chosen the voltage to be as low as possible while still giving a clear signal in order not to destroy the MCP by electrical breakdown, which could potentially suffer from debris due to its proximity to the interaction point.  Since the amplification factor of MCP-based systems could be sensitive to the signal level as well as duration, it is important to calibrate the detector in conditions as close as possible to the experimental conditions. We designed a pulsed calibration system to produce similar signals at the output of the TOF detector as were recorded during the experiment, both in voltage and in pulse width. 

The calibration system allows simultaneously recording signals from a ring-like Faraday cup and from our TOF detector positioned concentrically with the Faraday cup. The charge transmitted by the Faraday cup is calculated from first principles and is compared with the TOF signal, thus we can calculate the amplification factor of the TOF detector. 

The system is based on a HV Thyratron switch capable of switching $15$~kV at $\sim5$~ns. The negative pulse output is connected to a carbon brush which emits electrons at voltages $<-3$~kV. The radial homogeneity of charge area density was measured by using another Faraday cup, positioned concentrically with the ring-like Faraday cup instead of the TOF detector, thus comparing the signals of the two Faraday cups (accounting for their different areas). We measured higher charge density at the periphery compared to the center, which is expected since the outer ring-like Faraday cup is placed very close to the metal chamber. The radial distribution was accounted for in the calibration.

Electron pulses are used since they are much easier to be produced in the lab. Since the stopping range of both $15$~keV electrons and $100$~keV protons in the MCP is much smaller than the MCP thickness (smaller than $5\,\rm \mu m$ in both cases), the error introduced by using electrons instead of protons in the calibration process is negligible with respect to other sources. 

\begin{figure}[!h]
\includegraphics[width=0.7 \linewidth]{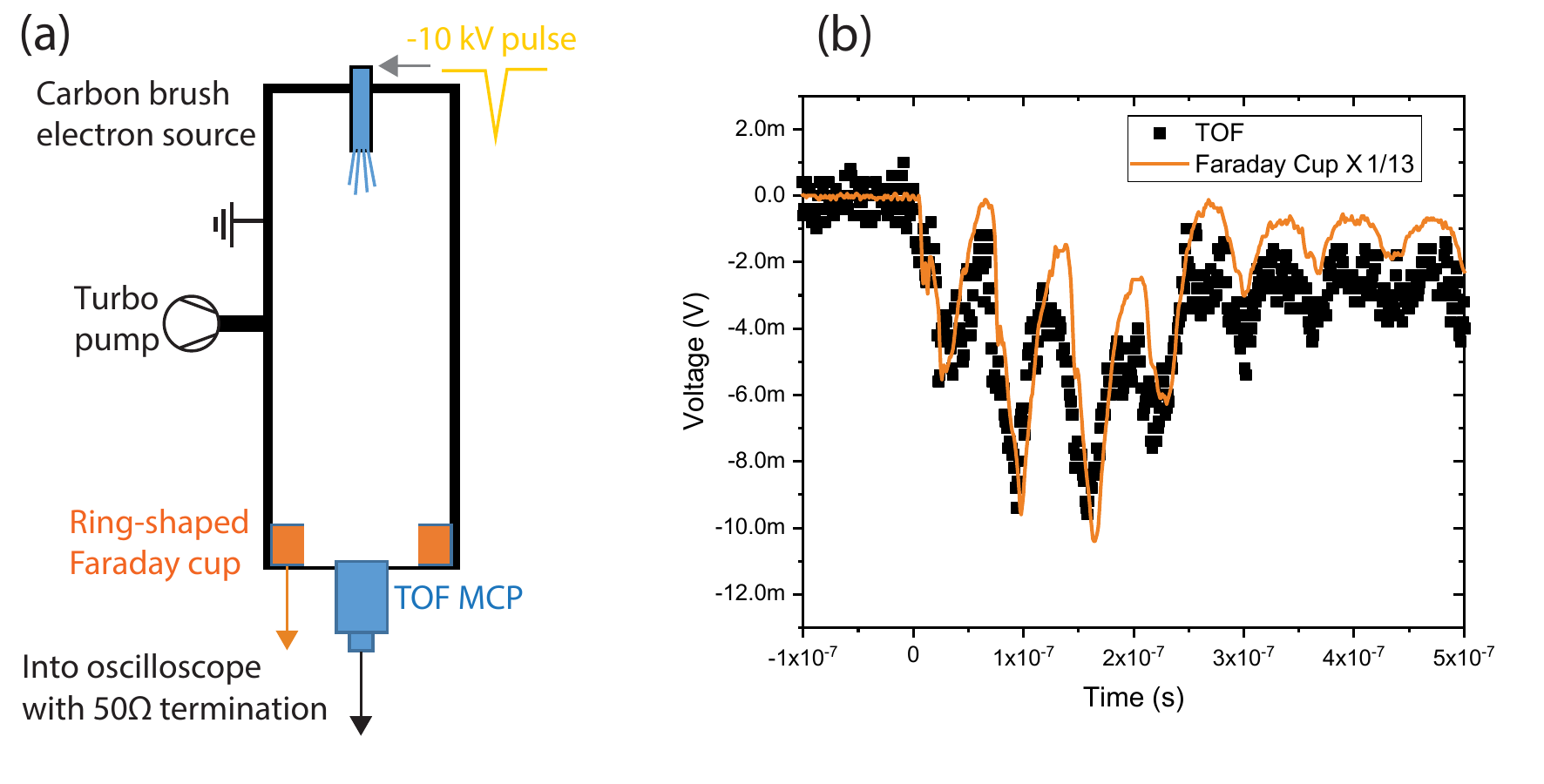}
\caption{(a) Schematic diagram of the calibration setup. (b) Simultaneous measurement of TOF and Faraday cup signals. In this shot the Faraday cup signal is divided by 13 to fit the TOF signal.}
\label{fig:TOF_calibration}
\end{figure}

\cref{fig:TOF_calibration} schematically shows the calibration setup alongside a simultaneous measurement of both TOF and Faraday cup signals. In order to estimate the MCP amplification factor, we divide the Faraday cup signal by some number so it fits best with the TOF signal. We estimate a $10\%$ fitting uncertainty in deriving this constant, which dominates the other possible sources of error. We find the MCP amplification factor to be $3\pm0.3$.

\subsection{TPS measurement}\label{sec:TPS_meas}

\begin{figure}[!h]
\includegraphics[width=0.4 \linewidth]{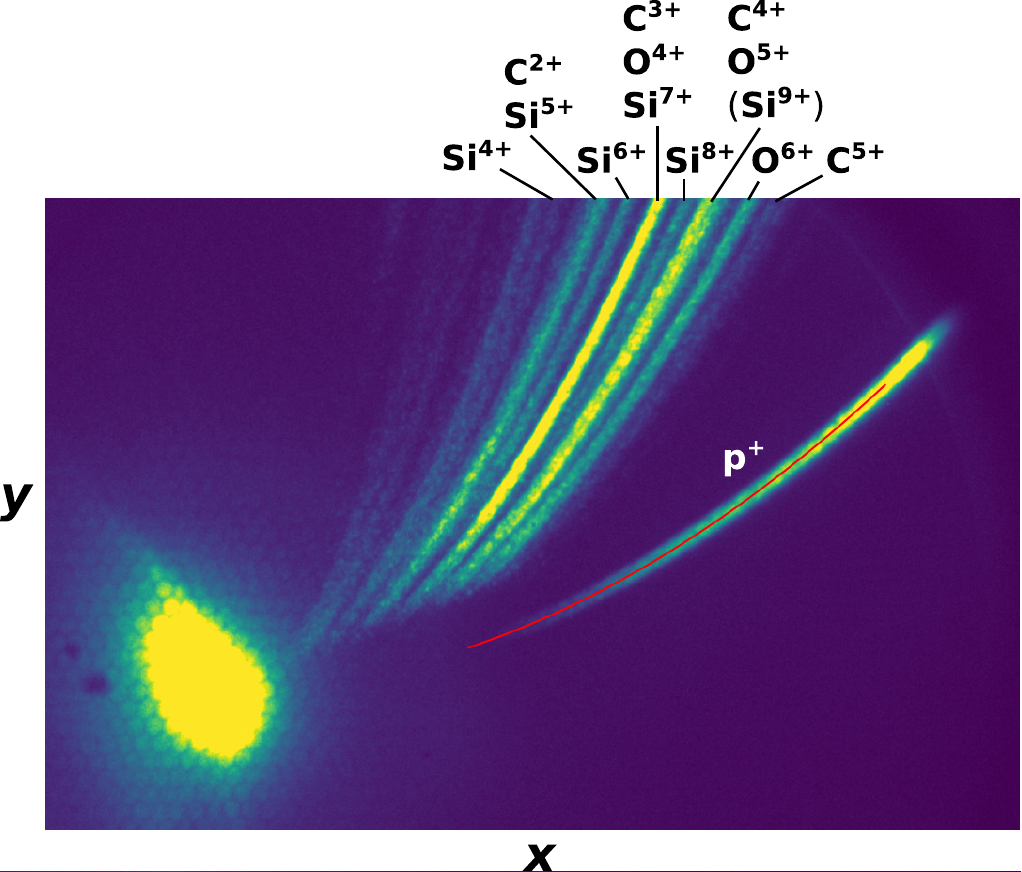}
\caption{A typical TPS measurement, showing the integrated signal of 100 consecutive shots.}
\label{fig:TPS_example}
\end{figure}

The Thomson parabola spectrometer is added as a small external chamber to the main chamber. The particles enter a $300\,\rm \mu m$ pinhole and are deflected by magnetic and electric fields. The magnetic field is produced by two permanent magnets $15$~mm long with a measured average of $0.32$~T over that length. The electric field is produced by two $50\times50\,\rm mm^2$ copper electrodes placed right after the magnets. The field strength is calculated to be $2$~kV/cm. The total distance from the far side of the magnets to the MCP is $210$~mm. The MCP is composed of two plates in a chevron setup.

In \cref{fig:TPS_example} we show an example of a TPS image. The red line follows the $H^{+}$ ions (protons) curve. The other usual contaminant ions species of carbon and oxygen are identified. Several curves correspond to charge-to-mass ratios of various silicon ions which are likely accelerated as well. 

The proton energy spectrum is obtained after averaging the values along the curve width, after subtraction of a mean background curve which is located in parallel to the signal curve below it. The background curve also has a width and both its mean and standard deviation are calculated along it. We define the cutoff energy as the lowest energy where the mean signal minus the standard deviation of the background is smaller than the mean background. The principal source of error in determining the maximum energy is geometric, i.e., the spatial extent of the ion source on the detector which manifests itself in the proton curve width. We find the relative error $E_{kin}/E_{kin}$ in our geometry to be $\Delta E_{kin}/E_{kin} \approx 0.072 (\frac{E_{kin}}{\text{MeV}})^{1/2}$.

\section{Analytic model}\label{Theory}

Let us consider plasma with a uniform electron density $n_{pe0}\gg n_c$ for $z<0$, and the evanescent preplasma ${n_{pe} = n_{pe0}\exp(-z/L_g)}$ for $z>0$. A p-polarized laser field impinges on plasma at an angle $\theta$ and reaches the depth where the electron density becomes critical for the laser $n_{pe}(z_c)\sim n_c$. For the very short gradients $L_g \ll \lambda$, the preplasma mainly reflects laser field but also develops such processes as resonant excitation of plasma electron modulations at $z\gtrsim z_c$, and the so-called Brunel absorption \cite{brunel_not-so-resonant_1987}. A more explicit description of this laser-plasma interaction could be developed with the help of numerical modelling \cite{veltcheva_brunel-dominated_2012}. There, the different trajectories of the surface electrons are numerically calculated and the resulting accelerating electric field is extracted from the resulting charge imbalance. Other than scaling laws with the laser duration and its amplitude, this approach however does not provide analytic expressions and does not incorporate the dependence on other interaction parameters such as the laser spot size and the preplasma gradient scale length.

For a qualitative theoretical description, let us simplify the picture by assuming the that laser removes electrons up to some depth $z_0\gtrsim z_c$, leaving a layer of un-neutralized heavy ions that we consider immobile during the whole process. The acceleration is produced by the electrostatic field of these ions and is thus directly related to the value of $z_0$. We further consider the electrons to be only displaced along the $z$-axis, which occurs as an instantaneous response to the laser field on the target. This assumption agrees with the experimental data and PIC simulations, where we see that $z_0$ is indeed a function of the laser intensity (see \cref{PIC}). The simplest way to correlate $z_0$ with the laser intensity is to consider that electron displacement is produced in a way to maintain static equilibrium between the radiation pressure of the incident and reflected laser field, ${P_\text{rad}=2\cos^2\theta\, I_\text{las}/c}$, and the electrostatic pressure $P_\text{es}$ generated by the charge separation. Qualitatively, the latter can be estimated from the capacitor model, $P_\text{es} \approx \sigma_i^2/ \epsilon_0$, with $\epsilon_0$ being vacuum permittivity and $\sigma_i= e\int_{z_0}^{\infty} n_{pe} dz$ is the area charge density of un-neutralized ions. Note that so far we have already discarded the ionization, heating and expansion of the heavy ion species, as well as the fast electrons ejected by the laser and present in the preplasma, ${z>z_0}$, (see \cref{fig:PIC_dynamics}(a)). With this, we may write the pressure balance condition as
\begin{equation}\label{Eq:BalanceFullField}
\sigma_i = \cos\theta\,\sqrt{2\epsilon_0\, I_\text{las} /c} = \epsilon_0 \hat{E}_\text{las} \cos\theta\,,
\end{equation}
where $\hat{E}_\text{las}=\sqrt{2 I_\text{las}/\epsilon_0 c}$ is the amplitude of laser electric field. For the considered preplasma profile this condition defines the penetration depth:
\begin{equation}\label{Eq:Z0FullField}
 z_0 = L_g \ln \left (\,\cfrac{e  n_{p0} L_g}{2\cos\theta} \sqrt{\frac{c}{\epsilon_0 I_\text{las}}} \,\right) = L_g \ln \left (\cfrac{2 \pi  \tilde{n}_{p0} \tilde{L}_g }{a_0 \cos\theta} \right)\,,
\end{equation}
where the dimensionless density $\tilde{n}_{p0}$ is in units of the critical plasma density $n_c = \pi/r_e\lambda^2$, $\tilde{L}_g$ is in units of $\lambda$, and $r_e$ is a classical electron radius (a similar result derived in a boosted frame can be found in \cite{vincenti_optical_2014}). Note that for a uniform plasma and normal laser incidence, this also leads to the well-known law of relativistic plasma transparency $a_0 = 2\pi \tilde{n}_{p0} \tilde{l}_0$, where $\tilde{l}_0$ is a laser penetration depth in units of $\lambda$ \cite{vshivkov_nonlinear_1998}. The obtained dependency \cref{Eq:Z0FullField} demonstrates a good agreement with the PIC simulations as presented in the inset of \cref{fig:PIC_dynamics}(b).

The resulting configuration can be presented as a positively charged plane (or disk in 3D) with density ${n_i = e n_{pe}(z>z_0)}$ attached to the surface of the solid neutral plasma. This description relies on the condition that the initial preplasma is very steep, $L_g\ll \lambda$, so that for $z\lesssim z_0$ the electron density is very high ($\gg n_c$) and the field of all electrons injected into the plasma by the laser is instantly screened. This phenomenon is an essential ingredient of the Brunel heating process, which allows electrons to carry away laser energy deep into the plasma, instead of building up a dense negative layer at the surface which is the case in longer preplasmas \cite{macchi_laser_2005-1, vincenti_optical_2014}. From this also follows that the electrostatic field of the ion layer, which is oriented inwards the plasma at $z\lesssim z_0$, is also effectively screened. This can be observed in \cref{fig:PIC_dynamics}(b), where one can see only the positive component of the $E_z$ calculated accounting for all the charges of the system. The accelerating, positive component of $E_z$ grows quickly with $z>z_0$, and reaches its maximum value, which we can estimate as ${E_{max} = \sigma_i/2\epsilon_0 = \frac{1}{2}\, \hat{E}_\text{las}\,\cos\theta}$. We can describe its spatio-temporal distribution of the accelerating field of the ion layer as:
\begin{equation}\label{Eq:Eacc}
 E_\text{acc}(t, z) =  \cfrac{\hat{E}_\text{las}(t)\,\cos\theta}{2} \; 
 \left[1-\exp\left(-\cfrac{z-z_0}{L_g}\right) \right]\;
 \left(1-\cfrac{z}{\sqrt{z^2+R_{i}^2}}\right)\,,
\end{equation}
where the size of the charged ion disk is determined by the projected laser spot size as $R_{i}=R_\text{las}/\cos\theta$. The second and third factors at the right hand side of \cref{Eq:Eacc} account for the field distribution within the preplasma and the three-dimensional geometry of the field, respectively.

\begin{figure}
\includegraphics[width=0.7\linewidth]{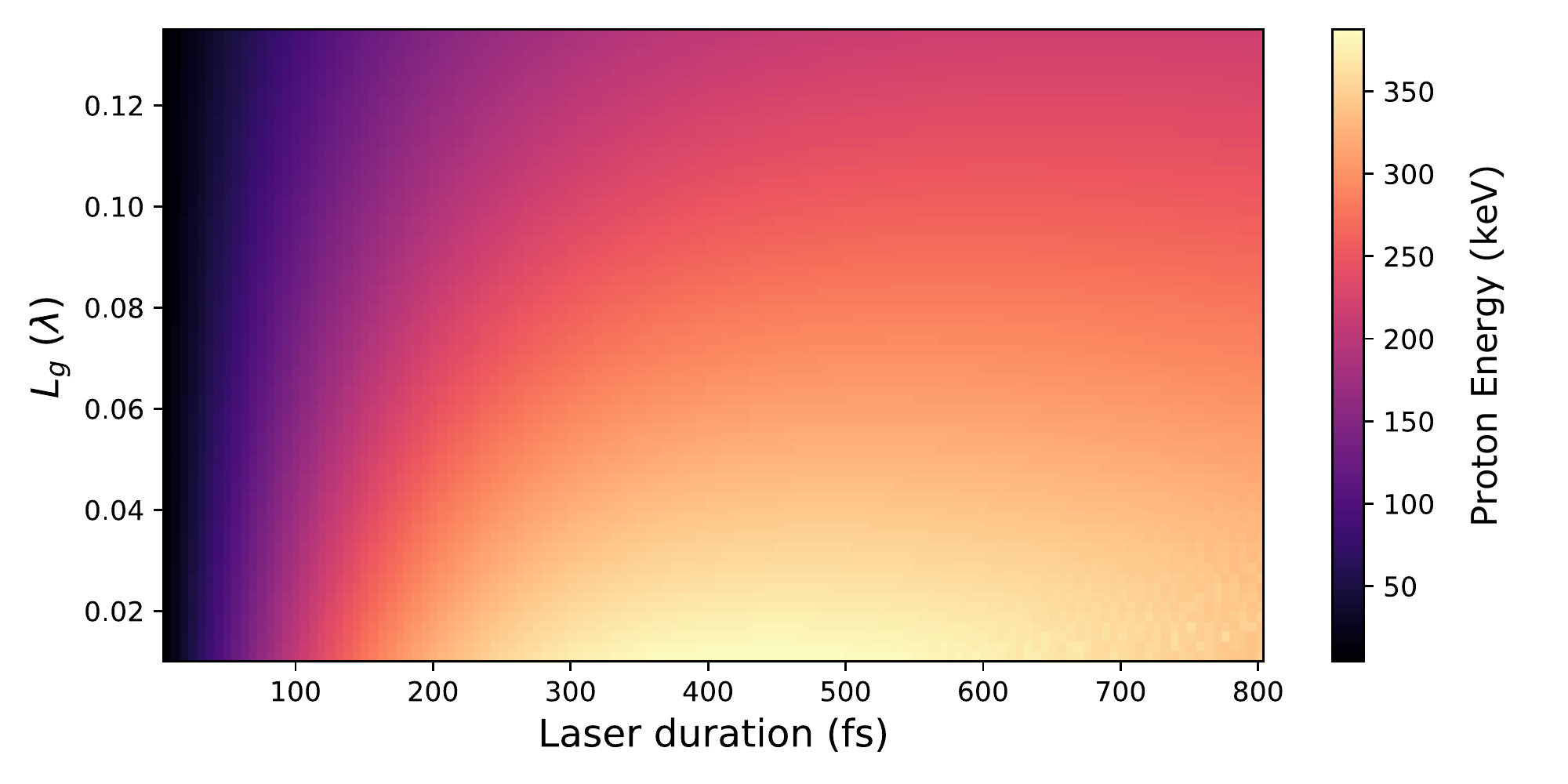}
\caption{Maximum proton energy in keV estimated from model \cref{Eq:Eacc,Eq:Motion} for different laser pulse durations and preplasma lengths.}
\label{fig:Toy_scan}
\end{figure}

Assuming that the accelerated protons remain non-relativistic, we may solve the equations of motion:
\begin{equation}\label{Eq:Motion}
 d_t p_p = e E_\text{acc}\,,\quad d_t z_p = p_p/m_p\,,
\end{equation}
and find the maximum gained energy, $W_\text{max} = p_\text{max}^2/2m_p$, where $p_\text{max}=e \int_{-\infty}^{\infty} E_\text{acc} \mathrm{d}t$ is the maximum proton momentum, and the integral is calculated along the proton trajectory $z_p(t)$. This integration can be done numerically, and for the parameters assumed for the simulations in \cref{PIC} ($W_\text{las}=1.5$~mJ, $R_\text{las}=1.8$~$\mu$m, $\theta=55\degree$ and $n_{p0} = 250\,n_c$), the model can describe the maximum gained energy as a function of the preplasma length and the laser pulse duration as presented in \cref{fig:Toy_scan}.

In the particular case of a very short laser pulse, the model \cref{Eq:Eacc,Eq:Motion} can be further simplified. Let us assume, that the total distance travelled by a proton during its acceleration, $L_\text{acc}\sim p_\text{max}/2m_p \tau_\text{las}$, is very small compared to the scale length of the field, $L_\text{acc}\ll R_{i}$, but is significantly longer than the preplasma gradient $L_\text{acc}\gg L_g$. With these conditions in mind,  we may neglect the last two factors in \cref{Eq:Eacc}, and estimate the energy gained by the protons in the uniform $E_{max}$ during the laser action. In this case, \cref{Eq:Motion} can be easily integrated giving:
\begin{equation}\label{Eq:LinScale}
 W_\text{max} = \sqrt{\frac{\ln 2}{\pi}}\; \cfrac{e^2 \cos^2\theta}{m_p c \,\epsilon_0}\;
\cfrac{W_\text{las} \,\tau_\text{las}}{R_\text{las}^2} \,.
\end{equation}

It is clear that \cref{Eq:LinScale} is only valid when $\tau \lesssim R_i/4\langle v_p\rangle$, where $\langle v_p\rangle=\sqrt{W_\text{max}/2m_p}$ is the proton velocity averaged over the acceleration. Assuming that this validity threshold  qualitatively determines the optimal acceleration regime, we may qualitatively correlate this with \cref{Eq:LinScale}, and find the scaling of the optimized acceleration as:
\begin{equation}\label{Eq:OptScale}
 W_\text{opt} = \left(\sqrt{\frac{2 \ln 2}{\pi m_p}} \cfrac{e^2}{4 c \,\epsilon_0} \cfrac{W_\text{las}}{R_i}\right)^{2/3}\,.
\end{equation}

\section{Particle-in-Cell modelling setup}\label{PIC_setup}

For insight into the acceleration mechanism we consider two-dimensional numerical modeling using the particle-in-cell (PIC) code WARPX \cite{vay_modeling_2021}. The numerical domain is defined as $x \in ( -52.5, 52.5)$~$\mu$m and $z \in (-32, 17)$~$\mu$m, and is resolved with the cells of $dx\times dz = 4\times 8.5$~nm size ($12288 \times 12288$ total cells). The plasma is presented by initially neutral but ionizable particles corresponding to the oxygen, silicon and hydrogen atoms. In the half-plane $z<0$, the matter is considered uniform, and for $z>0$ an exponential preplasma $\propto \exp(-|z|/L_g)$ is added. The considered material corresponds to fused silica with atomic number densities of particle species $n_\text{Si} = 2.19 \cdot 10^{22}$~cm$^{-3}$, $n_\text{O}=2\,n_\text{Si}$ in the uniform part. The hydrogen atoms origin from surface contaminants and are therefore only contained in the preplasma. As their real number density was not measured we choose to consider $n_\text{H} =  n_\text{Si} 10^{-6}$, which practically makes them trace particles with negligible self-fields. Within the layer which contains the preplasma and extends by a few skin-depths into plasma slab ($z>-1$~$\mu$m), all species are resolved with 16 particles per cell per species, and in the uniform plasma bulk the oxygen and silicon atoms are presented by 1 particle per cell per species. We have also included the pairwise Coulomb electron-ion collision for all ion species.

In the simulations, the laser beam has Gaussian temporal and spatial profiles, and is focused into a spot of $R_\text{las} = 1.8$~$\mu$m FWHM obliquely on the plasma surface with an angle of $55^\circ$. In the experiment, the longer pulses, $\tau_\text{las}>27$~fs, were produced by adding a linear chirp, but in the simulations we model all pulses as Fourier transform limited. We have validated this simplification using test runs with a chirped laser which have not indicated any significant difference with the FTL pulses of the same duration. 

\begin{figure*}
\includegraphics[width=0.9\linewidth]{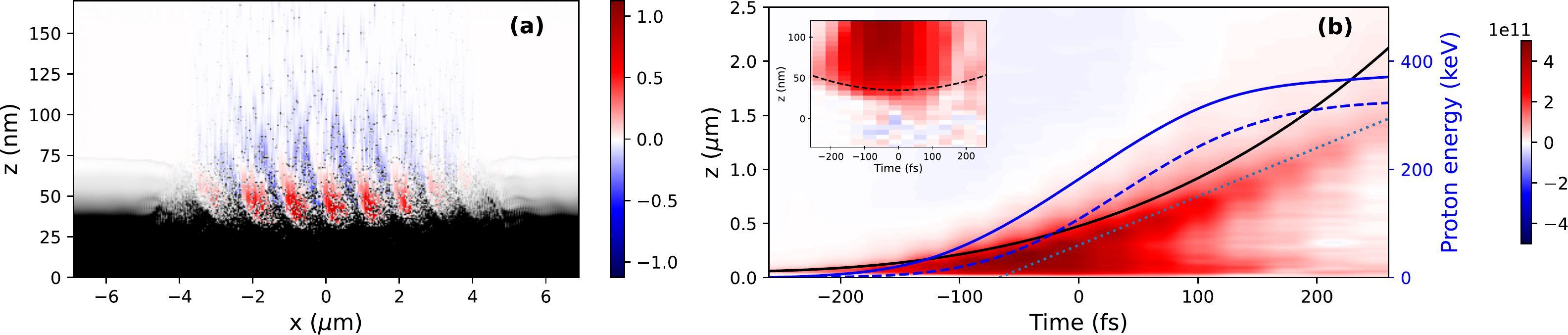}
\caption{(a) Charge density (red-blue, $n_c$ units) and electron density (grey, normalized for visibility) for $\tau_{las}=200$~fs pulse at $t=-240$~fs before peak field arrival; (b) Temporal evolution of $x$-averaged electrostatic field $\langle E_z\rangle$ (red-blue colormap), maximum proton energy (blue curve) and corresponding $z$-coordinate (black curve), maximum proton energy from \cref{Eq:Z0FullField,Eq:Eacc} (blue dashed curve); (b):inset Zoomed $\langle E_z\rangle$ (colors) and depth estimate \cref{Eq:Z0FullField} (black dashed curve);}
\label{fig:PIC_dynamics}
\end{figure*}

\section{Analysis of a specific simulation case}\label{PIC}

For a demonstration case that highlights the involved physical processes, we consider a simulation with the same plasma profile as in \cref{Theory,PIC_setup}, and the laser with $1.8$~mJ energy and FWHM duration ${\tau_\text{las}=200}$~fs. In this simulation, the numerical domain has the size $40\times 48$~$\mu$m, and is resolved with square cells of $6$~nm size; the plasma consists of silicon ions pre-ionized to $Z_\text{Si}=8$ and electrons with the density $n_{pe0} = 250\, n_c$, where $n_c=1.742\cdot 10^{21}$~cm$^{-3}$ being the critical plasma density for $\lambda=800$~nm. The plasma kinetics were modelled using $64$ macro-particles per cell per species in the interaction domain, $z>-140$~nm, while the ``deeper'' plasma bulk was resolved with a single macro-particle per cell per species. For this case no ionization nor electron-ion collisions were considered.

\Cref{fig:PIC_dynamics}(a) presents the charge and electron density maps the interaction region where acceleration occurs, and at the early moment ${t=-240}$~fs before the peak field arrival ($t=0$). One may see the none-neutral layer, where electrons are expelled by the laser, which also modulates plasma with modulations reaching down to $z\approx -34$~nm. The generated charge density reaches $\rho \approx +e n_c$ (red colors in \cref{fig:PIC_dynamics}(a)), which roughly agrees with the theoretical predictions of interaction depth of $z\approx -50$~nm.

The field that accelerates protons is generated by the charge density and \Cref{fig:PIC_dynamics}(a) and in simulation it is typically overlapped with the electromagnetic wave, so it cannot be directly analysed. In order to evaluate this field we reconstruct it from the instantaneous charge density distribution obtained from the PIC simulation. For this we consider the Poisson equation:
\begin{equation}\label{poisson_eq}
 \nabla^2 \phi = - \rho/\epsilon_0\,,\quad \mathbf{E} = -\nabla\phi\,,
\end{equation}
and assuming periodic boundaries we solve \cref{poisson_eq} using 2D Fourier transforms:
\begin{equation}\label{poisson_eq_sol}
 \mathbf{E} =  \mathrm{Re}\left(\mathrm{FFT}^{(-1)}\left[ \cfrac{-i\;\mathbf{k} \;\mathrm{FFT}[\rho]}{\epsilon_0\, k^2} \right]  \right)\,,
\end{equation}
where $\mathrm{FFT}$ and $\mathrm{FFT}^{(-1)}$ are the forward and inverse  Fourier transform operators and $k$ is the wave-vector.

The details of the simulated interaction dynamics are summarized in \cref{fig:PIC_dynamics}(b). The accelerating field calculated from \cref{poisson_eq,poisson_eq_sol} and averaged as $\langle E_z \rangle = 1/(2 x_0)\int_{-x_0}^{x_0} E_z \mathrm{d}x$ with $x_0=1.5\,\rm \mu m$,  is plotted in red-blue colors. This field appears at the time of laser arrival near $z_0$, that agrees with the estimate from \cref{Eq:Z0FullField} (dashed black curve in the inset). For $z>z_0$, this field quickly grows and forms a broad plateau region, after which it diminishes and vanishes at $z\gtrsim 0.5$~$\mu$m. This profile remains almost unchanged for $t<0$.

The evolution of the maximum proton energy and corresponding $z$-coordinate are shown in \cref{fig:PIC_dynamics}(b) with blue and black solid curves, respectively. Considering the motion of a test-proton in the theoretical field we can also find the energy evolution that agrees well with the simulation (blue dashed curve). At the later times, the field starts to expand towards vacuum with the velocity ${v_\text{exp}\approx 4.5\times 10^6}$~m/s  (dotted line). This process is driven by the expansion of the Si preplasma layer, and for the protons that propagate in phase with this layer this allows gaining further energy. At the same time, the field of an expanding layer acquires transverse components due to its curvature, which broadens the proton beam divergence. 

\section{Indications of TNSA effect on accelerated protons}\label{TOF_mixing}

\begin{figure*}
\includegraphics[width=0.9\linewidth]{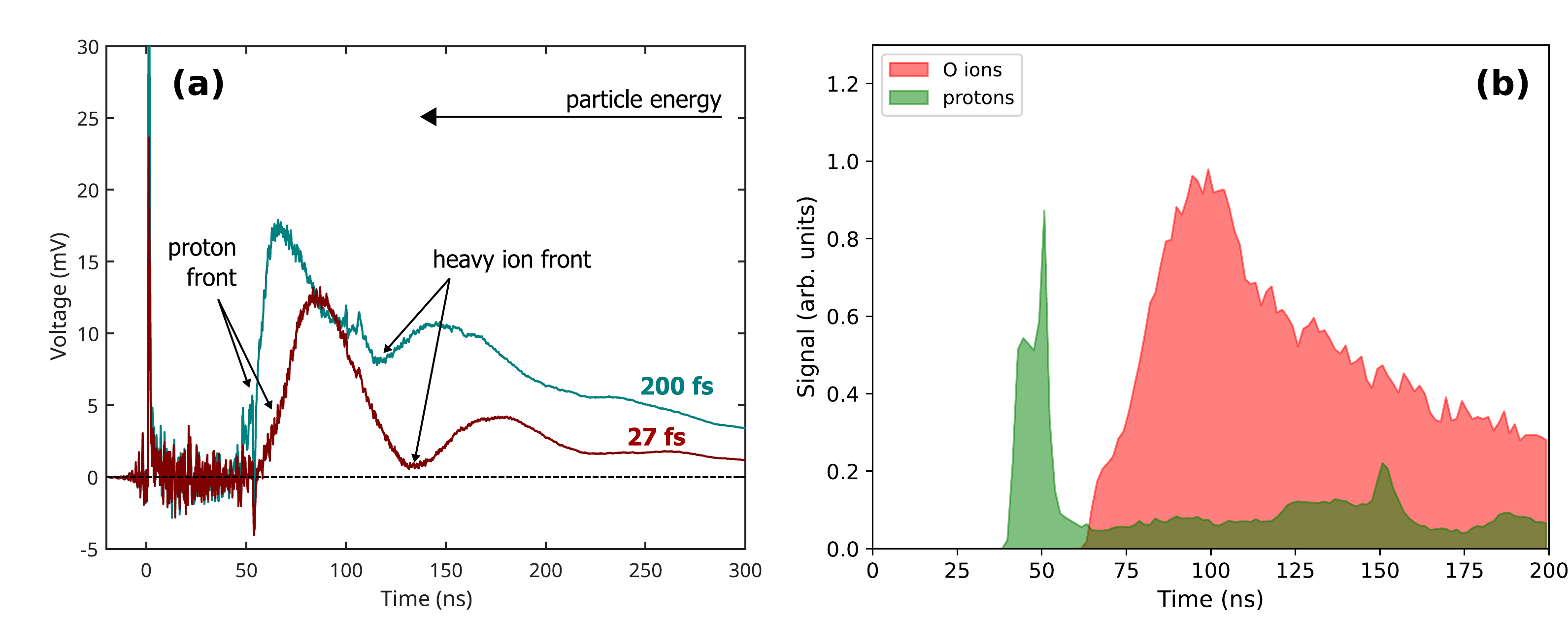}
\caption{(a) TOF measurements for 27fs and 250 fs laser durations (with voltage multiplied by -1) (b) Simulated TOF measurement for 100 fs laser pulse and pinhole with angular aperture of $0.8\degree$.}
\label{fig:TOF_mixing}
\end{figure*}

The effect of TNSA ions on the protons is mainly observed in the simulations, but its signatures can also be seen in the TOF measurements. As one can see \cref{fig:TOF_mixing}(a) (and in \cref{fig:TOF_signal_spect}), the TOF detector signal contains two distinctive peaks corresponding to the accelerated protons and heavy ions respectively. For the short laser pulses (red curve in \cref{fig:TOF_mixing}(a)) the signal between the peaks falls down to zero, which suggests spatial separation of the species. For the long pulses, the signal between the peaks becomes very significant, indicating the overlap of ions and lower energy protons. 

A very similar picture can be obtained from the PIC simulations, and it is shown for the case of a 100 fs laser in \cref{fig:TOF_mixing}(b). For this figure, we have considered the velocities of protons and oxygen ions at the end of the acceleration, and have translated them to the time delays $t_\text{TOF} = L_\text{TOF} / v_z$, where $L_\text{TOF} = 365$~mm is the distance between the target and TOF detector in the experiment. We have also made a selection of the particles corresponding to a pinhole with a full angular aperture of $0.8\degree$, which is centered on the direction of the proton bunch. Since in the simulations we set up the proton layer with a very low density (test particles), the curves corresponding to the ions (red) and protons (green) are normalized individually. From \cref{fig:TOF_mixing}(b) is easy to see that the proton signal falls abruptly in the time-range where ions and protons overlap, which corresponds to the degradation of angular divergence observed in Fig.~3(c) for this case.

\bibliography{bibliography.bib}
\bibliographystyle{apsrev4-1}